\documentclass[12pt]{amsart}

\usepackage{amsmath}
\usepackage{amssymb}
\usepackage{epic}
\usepackage[final]{graphicx}

\newtheorem{thm}{Theorem}%[section]
\theoremstyle{definition}

\theoremstyle{remark}
\newtheorem{remark}{Remark}[thm] % \renewcommand{\theremark}{}
\newtheorem{ex}{Example}[thm] 
\theoremstyle{plain}

\numberwithin{equation}{section}

\def\NN{{\mathbb N}}

\def\RR{{\mathbb R}}

\def\NDi{N_{\operatorname{D}}}
\def\NNe{N_{\operatorname{N}}}

\def\Area{\operatorname{Area}}

\def\scrA{{\mathcal A}}
\def\scrD{{\mathcal D}}

\def\scrK{{\mathcal K}}

\begin{document}

\title{Quantum leaks}
\author{Jens Marklof}
\address{School of Mathematics, University of Bristol,
Bristol BS8 1TW, U.K.} 
\email{j.marklof@bristol.ac.uk}
\date{\today}

\thanks{Research supported by an EPSRC Advanced Research Fellowship and EPSRC Research Grant GR/T28058/01.}

\begin{abstract}
We show that eigenfunctions of the Laplacian on certain non-compact domains with finite area may localize at infinity---provided there is no extreme level clustering---and thus rule out quantum unique ergodicity for such systems. The construction is elementary and based on `bouncing ball' quasimodes whose discrepancy is proved to be significantly smaller than the mean level spacing.
\end{abstract}

\maketitle

\section{Introduction}

Consider a region $\scrD$ in $\RR^2$ with piecewise smooth boundary and finite area. The {\em billiard flow} on the unit cotangent bundle of $\scrD$ is defined as the motion along straight lines with specular reflections at its boundary $\partial\scrD$. The quantum states and energy levels of the flow are determined by the eigenvalue problem for the Dirichlet Laplacian,\footnote{Our results can easily be adapted to the case of Neumann boundary conditions provided the spectrum of the Laplacian is discrete (which, in contrast to Dirichlet conditions, is not generally the case for non-compact regions with finite area).}
\begin{equation}
\begin{cases}
(\Delta+\lambda)\varphi =0 \\
\varphi\big|_{\partial\scrD} =0 ,
\end{cases}
\end{equation}
where $\Delta=\partial_x^2+\partial_y^2$. It is well known that the spectrum is discrete. The asymptotic distribution of the eigenvalues 
\begin{equation}
0<\lambda_1\leq\lambda_2\leq\ldots\to\infty 
\end{equation}
is governed by Weyl's law (cf. \cite{Simon79,Berg92a,Berg92b,Ivrii98,Berg01} and references therein)
\begin{equation}\label{weyl}
\lim_{\lambda\to\infty}\frac{\#\{ j : \lambda_j < \lambda \}}{\lambda} = \frac{\Area(\scrD)}{4\pi}.
\end{equation}
The mean spacing between consecutive eigenvalues is therefore asymptotically constant. We denote by $\{\varphi_j\}_j$ an orthonormal basis of eigenfunctions, and consider the probability measure
\begin{equation}
d\nu_j = |\varphi_j(x,y)|^2 dx\,dy
\end{equation}
associated with the $j$th eigenstate. One of the central problems in quantum chaos is to classify all weak limits of $d\nu_j$ as $j\to\infty$. The {\em quantum ergodicity theorem}, due to Schnirelman, Zelditch and Colin de Verdi\`ere \cite{Schnirelman74,Zelditch87,Colin85} (adapted for billiard flows on domains of the above type in \cite{Zelditch96}), asserts that, if the underlying dynamics is ergodic, there is a subsequence $\lambda_{j_1},\lambda_{j_2},\ldots$ of full density\footnote{A subsequence $\{\lambda_{j_i}\}_i$ is of full density if $\lim_{\lambda\to\infty} \#\{ i : \lambda_{j_i} < \lambda \}/\#\{ j : \lambda_{j} < \lambda \} = 1$.}
such that the corresponding eigenfunctions $\varphi_{j_i}$ $(i\to\infty)$ become uniformly distributed on the unit cotangent bundle of $\scrD$. This implies for instance that for any set $\scrA\subset\scrD$ with smooth boundary,
\begin{equation}
	\lim_{i\to\infty} \int_{\scrA} d\nu_{j_i} = \frac{\Area(\scrA)}{\Area(\scrD)}.
\end{equation}
The proof of this theorem does not indicate whether in fact {\em all} eigenfunctions become uniformly distributed (a phenomenon called {\em quantum unique ergodicity} since there is only one possible quantum limit \cite{Rudnick94,Sarnak03}), or if there may exist sparse subsequences that have a singular limit, e.g., measures concentrated on periodic orbits of the billiard flow. Such exceptional subsequences have been observed in numerical experiments and are referred to as {\em scars} or {\em bouncing ball modes}. Following earlier results for quantum maps \cite{Degli95,Marklof00,Kurlberg00,Kurlberg01}, recent seminal contributions on the question of quantum unique ergodicity include the work of Faure, Nonnenmacher and De Bi\`evre \cite{Faure03,Faure04} who prove the existence of localized eigenstates for quantum cat maps, and Lindenstrauss' proof \cite{Lindenstrauss03} of quantum unique ergodicity in the case of Hecke eigenstates\footnote{Hecke eigenstates are simultaneous eigenfunctions of the Laplacian and all Hecke operators. If the spectrum of the Laplacian is simple, as conjectured e.g. for the modular surface, any eigenfunction of the Laplacian is a Hecke eigenstate.} of the Laplacian on compact arithmetic hyperbolic surfaces of congruence type.

\setlength{\unitlength}{0.00008\textwidth}
\begin{figure}
{\renewcommand{\dashlinestretch}{30}
\begin{picture}(8774,1227)(0,-10)
\drawline(1175,1200)(2137,1200)(2137,612)
	(3500,612)(3500,187)(5687,187)
	(5687,62)(8762,62)
\drawline(8762,12)(12,12)
\drawline(12.000,25.000)(85.790,43.131)(158.583,64.924)
	(230.196,90.326)(300.451,119.272)(369.173,151.691)
	(436.190,187.502)(501.335,226.615)(564.445,268.933)
	(625.363,314.350)(683.937,362.753)(740.020,414.020)
	(793.473,468.025)(844.162,524.631)(891.961,583.699)
	(936.750,645.080)(978.418,708.621)(1016.860,774.164)
	(1051.981,841.545)(1083.693,910.597)(1111.917,981.145)
	(1136.582,1053.015)(1157.627,1126.028)(1174.999,1200.000)
\end{picture}
}
\caption{Leaky Sinai billiard} \label{fig1}
\end{figure}
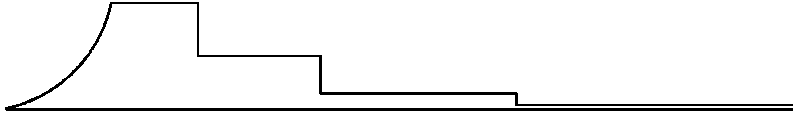
\begin{figure}
{\renewcommand{\dashlinestretch}{30}
\begin{picture}(8784,1235)(0,-10)
\drawline(1185,1208)(2147,1208)(2147,620)
	(3510,620)(3510,195)(5697,195)
	(5697,70)(8772,70)
\drawline(8772,20)(22,20)
\drawline(1197.000,1208.000)(1121.977,1205.533)(1047.258,1198.336)
	(973.144,1186.436)(899.928,1169.882)(827.903,1148.739)
	(757.356,1123.092)(688.569,1093.043)(621.816,1058.712)
	(557.363,1020.236)(495.467,977.768)(436.376,931.478)
	(380.324,881.550)(327.536,828.183)(278.221,771.591)
	(232.577,711.999)(190.786,649.644)(153.014,584.776)
	(119.412,517.653)(90.114,448.542)(65.237,377.721)
	(44.880,305.470)(29.124,232.078)(18.033,157.838)
	(11.649,83.046)(10.000,8.000)
\end{picture}
}
\caption{Leaky Bunimovich billiard} \label{fig2}
\end{figure}
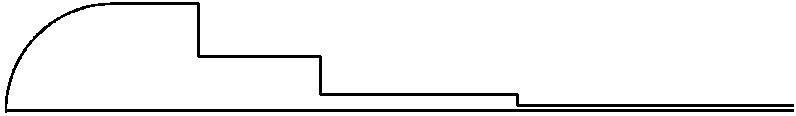
\begin{figure}
{\renewcommand{\dashlinestretch}{30}
\begin{picture}(8774,1227)(0,-10)
\drawline(8762,12)(12,12)
\drawline(1200,1200)(12,1200)(12,25)
\drawline(1175,1200)(2137,1200)(2137,612)
	(3500,612)(3500,187)(5687,187)
	(5687,62)(8762,62)%(8750,100)
\end{picture}
}
\caption{Leaky polygonal billiard} \label{fig3}
\end{figure}
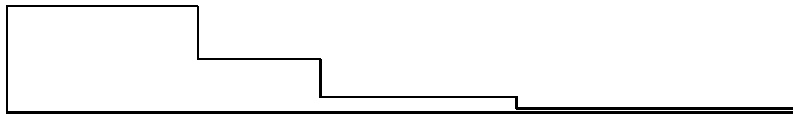

In the present paper we show that for certain non-compact domains $\scrD\subset\RR^2$ with finite area the sequence of measures $d\nu_j$ is not tight,\footnote{A sequence of probability measures $d\nu_j$ is {\em tight} if for any $\epsilon>0$ there is a compact domain $\scrK\subset\scrD$ such that $\limsup_{j\to\infty}	\int_{\scrD-\scrK} d\nu_{j} < \epsilon$.} provided there is no extreme clustering of eigenvalues. Hence there exist subsequences of eigenstates $\varphi_{j_i}$ that leak to infinity, and quantum unique ergodicity is not satisfied for such a system. 

Let $\scrD$ be given by 
\begin{equation}
\scrD=\{ (x,y)\in\RR^2 : x> 0,\; 0< y < f(x)\}
\end{equation}
where $f:(0,\infty)\to (0,\infty)$ is right-continuous and decreasing to $0$ as $x\to\infty$. More specifically, we assume that $f$ is constant on the intervals $[a_i,a_{i+1})$, $i=1,2,3,\ldots$. Examples of such domains are displayed in figs. \ref{fig1}--\ref{fig3}. The condition
\begin{equation}
\sum_{i=1}^\infty \ell_i\delta_i < \infty, \qquad \text{with $\delta_i:=f(a_i)$ and $\ell_i:=a_{i+1}-a_i$,}
\end{equation}
ensures $\scrD$ has finite area. 
To illustrate our main result, let us for example choose $\delta_i=i^{-(1+\sigma)}$ and $\ell_i=i^\rho$ where $\sigma>\rho>0$ are abribrary fixed constants. Theorem \ref{thm1} in Section \ref{secLeaky} implies that there is a constant $C>0$ such that (at least) one of the following two statements is true:
\begin{itemize}

\item[$\Box$] There is a subsequence of eigenfunctions $\varphi_{j_i}$ ($i=1,2,\ldots$) with eigenvalues $\lambda_{j_i}\in\pi^2 i^{2(1+\sigma)}+[-C i^{-2\rho},C i^{-2\rho}]$ and some $c>0$ such that for any compact $\scrK\subset\scrD$ we have
\begin{equation}
\liminf_{i\to\infty} \int_{\scrD-\scrK} d\nu_{j_i} >c.
\end{equation}	

\item[$\Box$] The number of eigenvalues $\lambda_j$ in the interval $\pi^2 i^{2(1+\sigma)}+[-C i^{-2\rho},C i^{-2\rho}]$ is unbounded as $i\to\infty$.

\end{itemize}
The first statement implies that eigenfunctions loose a positive proportion of mass. The second alternative implies extreme level clustering; this seems unlikely for a generic billiard of the above type, but cannot a priori be ruled out. To get a rough idea on whether to expect more level clustering than in the case of compact domains $\scrD$, we show in Section \ref{secThm2} that the spectral counting function has the asymptotics (Theorem \ref{thm2})
\begin{equation}
\#\{ j : \lambda_j < \lambda \} = \frac{\Area(\scrD)}{4\pi}\,\lambda - \frac{L(\lambda)}{4\pi} \sqrt\lambda + \frac{1}{2\pi} \sqrt\lambda
\sum_{\substack{i=1\\ \delta_i\sqrt\lambda>\pi}}^\infty \ell_i  \sum_{r=1}^\infty \frac1r J_1\bigg(2 r \delta_i \sqrt{\lambda}\bigg) +O(\sqrt\lambda),
\end{equation}
where 
\begin{equation}
L(\lambda) = 2 \sum_{\substack{i=1\\ \delta_i\sqrt\lambda>\pi}}^\infty \ell_i 
\end{equation}
is an `effective length' of the boundary $\partial\scrD$ and $J_1$ is the $J$-Bessel function. The fluctuations are therefore larger than in the compact case, where the error term is of order $O(\sqrt\lambda)$; cf. Section \ref{secThm2} for a more detailed discussion.

The proof of Theorem \ref{thm1} is elementary and based on the construction of `bouncing ball' quasimodes \cite{Heller88,Backer97,Tanner97,Donnelly03,Burq04,Burq03a,Burq03b,Zelditch04,Hillairet05} (see also Bogomolny and Schmit's recent work on eigenfunctions in pseudo-integrable billiards \cite{Bogomolny04}). The non-compactness of the domain allows for quasimodes with discrepancy almost as small as $O(\mu^{-1})$, where $\mu$ is the quasi-eigenvalue. The best rigorous bound for the discrepancy in the compact case is $O(1)$, cf. \cite{Donnelly03}.

Our construction is completely independent on the choice of $f$ on the interval $(0,a_1)$, and one may use this additional freedom to tune $f$ on $(0,a_1)$ in such a way that the billiard flow on $\scrD$ is ergodic. It seems plausible that this is the case if the billiard flow on the restricted compact region $\scrD_0=\{ (x,y)\in\RR^2 : 0<x<a_1,\; 0< y < f(x)\}$ is ergodic (as in the examples displayed in figs. \ref{fig1} and \ref{fig2}), but to the best of my knowledge there are no rigorous results in this direction (see however \cite{Lenci02,Lenci03,Graffi04} for proofs of ergodicity for different classes of non-compact domains). A further interesting class of examples are infinite pseudo-integrable billiards (fig. \ref{fig3}) that are known to be ergodic\footnote{Since the modulus of the momentum components in both $x$- and $y$-directions are constants of motion, ergodicity is here understood with respect to a two-dimensional submanifold of the unit cotangent bundle.} for almost all initial directions \cite{Degli00}.

\section{Quasimodes}

A function $\psi\in H_0^1(\scrD)$ is called a {\em quasimode} for $-\Delta$ with {\em quasi-eigenvalue $\mu$} and {\em discrepancy $\epsilon$}, if
\begin{equation}\label{qmode}
\begin{cases}
\| (\Delta+\mu)\psi \| \leq \epsilon \|\psi\| , \\
\psi\big|_{\partial\scrD} = 0,
\end{cases}
\end{equation}
where $\|\,\cdot\,\|$ denotes the $L^2$ norm.
A sequence of quasimodes $\{\psi_i\}_i$ with quasi-eigenvalues $\mu_i$ {\em is of order $s$}, if 
\begin{equation}\label{qmode2}
\| (\Delta+\mu_i)\psi_i \| = O(\mu_i^{-s/2}) \|\psi_i\| .
\end{equation}
We summarize a few important properties of quasimodes; more details can be found in \cite{Colin77,Lazutkin93,Donnelly03,Zelditch04}.

By expanding $\psi$ in an orthonormal basis of eigenfunctions, $\psi=\sum_j \langle \psi,\varphi_j \rangle \varphi_j$, it is easy to see that \eqref{qmode} implies
\begin{equation}
\sum_j |\langle \psi,\varphi_j \rangle|^2 (\lambda_j-\mu)^2 \leq \epsilon^2 \|\psi\|^2
=
\epsilon^2 \sum_j |\langle \psi,\varphi_j \rangle|^2.
\end{equation}
Hence $|\lambda_j-\mu|\leq \epsilon$ for at least one $j$, i.e., there is at least one eigenvalue $\lambda_j$ in the interval $[\mu-\epsilon,\mu+\epsilon]$. Consider the larger interval $J=[\mu-b\epsilon,\mu+b\epsilon]$, $b>1$. We have
\begin{equation}\label{ring}
\sum_{\lambda_j \notin J} |\langle \psi,\varphi_j \rangle|^2 
\leq (b\epsilon)^{-2} \sum_{\lambda_j \notin J} |\langle \psi,\varphi_j \rangle|^2 (\lambda_j-\mu)^2  
\leq b^{-2} \|\psi\|^2 .
\end{equation}
For a domain $\scrA\subset\scrD$ define
\begin{equation}
\| \psi \|_\scrA= \sqrt{\int_{\scrA} |\psi(x,y)|^2 dx\,dy} .
\end{equation}
Triangle and Cauchy-Schwarz inequality imply
\begin{equation}
\begin{split}
\| \psi \|_\scrA 
& \leq \bigg\|\sum_{\lambda_j\in J} \langle\psi,\varphi_j\rangle \varphi_j\bigg\|_\scrA  + 
\bigg\|\sum_{\lambda_j\notin J} \langle\psi,\varphi_j\rangle \varphi_j\bigg\|_\scrA \\
& \leq \sqrt{\sum_{\lambda_j\in J} |\langle\psi,\varphi_j\rangle|^2} \sqrt{\sum_{\lambda_j\in J} \|\varphi_j\|_\scrA^2} + 
\bigg\|\sum_{\lambda_j\notin J} \langle\psi,\varphi_j\rangle \varphi_j\bigg\| \\
& \leq \|\psi\| \sqrt{\sum_{\lambda_j\in J} \|\varphi_j\|_\scrA^2} + \sqrt{\sum_{\lambda_j\notin J} |\langle\psi,\varphi_j\rangle|^2 }
\end{split}
\end{equation}
and hence, together with \eqref{ring},
\begin{equation}
\sqrt{\sum_{\lambda_j\in J} \|\varphi_j\|_\scrA^2} \geq
\frac{\| \psi \|_\scrA}{\|\psi\|} - b^{-1} .
\end{equation}
Now suppose that 
\begin{equation}\label{assu}
\text{\begin{minipage}{0.8\columnwidth}{\em for a sequence of quasimodes $\psi_i$ with quasi-eigenvalue $\mu_i$ and discrepancy $\epsilon_i$ the intervals $J_i=[\mu_i-b\epsilon_i,\mu_i+b\epsilon_i]$ each contain at most $k$ eigenvalues $\lambda_j$.}\end{minipage}}
\end{equation}
Then, in each interval $J_i$ there is a $\lambda_{j_i}$ such that
\begin{equation}\label{tru}
\|\varphi_{j_i}\|_\scrA \geq \frac{1}{\sqrt k}\bigg(
\frac{\| \psi_i \|_\scrA}{\|\psi_i\|} - b^{-1} \bigg).
\end{equation}

\section{Leaky domains\label{secLeaky}}

Let $f:(0,\infty)\to (0,\infty)$ be a right-continuous function, monotonically decreasing to 0 on the half-line $[a_1,\infty)$ (for some $a_1>0$), and $\int f(x)dx < \infty$. We are interested in the domain $\scrD=\{ (x,y)\in\RR^2 : x> 0,\; 0< y < f(x)\}$. In the following we will assume that $f$ is chosen so that 
\begin{equation}\label{fina}
\int_{a_1}^\infty f(x) h(\pi^2 f(x)^{-2}) dx < \infty, 
\end{equation}
where $h:[0,\infty)\to[0,\infty)$ is a fixed increasing function bounded by $h(x)\leq\sqrt x$. 
The central result is the following.\footnote{The notation $A\ll B$ for two positive quantities $A,B$ means {\em there is a constant $C>0$ such that $A\leq C B$}. We write $A\asymp B$ if $A\ll B \ll A$.} 

\begin{thm}\label{thm1}
For any given decreasing function $\tau:[0,\infty)\to(0,\infty)$, and any infinite sequence of real numbers
\begin{equation}\label{seq}
0<\mu_1 \leq \mu_2 \leq \ldots \to \infty
\end{equation}
satisfying
\begin{equation}\label{one}
\sum_{i=1}^\infty \tau(\mu_i) < \infty,
\end{equation}
there is a domain $\scrD$ of the above type whose Dirichlet Laplacian has an infinite sequence of quasimodes $\psi_{i,m,n}$ with quasi-eigenvalues 
\begin{equation}\label{quas}
\mu_{i,m,n}= n^2 \mu_i + m^2 \xi_i, \qquad i,m,n\in\NN, 
\end{equation}
and
\begin{equation} \label{eps}
\xi_i \asymp \frac{h(\mu_i)^2}{\mu_i\, \tau(\mu_i)^2},
\end{equation}
so that
\begin{itemize}
\item[(i)]
$\| (\Delta+\mu_{i,m,n}) \psi_{i,m,n} \| = O(m\xi_i) 
\| \psi_{i,m,n} \|$,
\item[(ii)]
$\langle \psi_{i,m,n}, \psi_{i',m',n'} \rangle  = 0$ for $i\neq i'$ or $n\neq n'$, 
\item[(iii)]
$|\langle \psi_{i,m,n}, \psi_{i,m',n} \rangle |\ll \min\{0.001, |m-m'|^{-1}\} \|\psi_{i,m,n}\|\,\|\psi_{i,m',n}\|$ for $m\neq m'$,
\item[(iv)]
for any compact set $\scrK\subset\scrD$,
\begin{equation}\label{three}
\frac{\|\psi_{i,m,n}\|_{\scrD-\scrK}}{\|\psi_{i,m,n}\|} \to 1 
\end{equation}
uniformly for all $m,n\in\NN$ as $i\to\infty$.
\end{itemize}
\end{thm}

\begin{remark}
Note that the set $\{\mu_{i,m,n}: i,m,n\in\NN\}$ is a discrete subset of $\RR_+$, with mean density
\begin{equation}\label{weyl2}
\lim_{\lambda\to\infty}\frac{\#\{ (i,m,n) : \mu_{i,m,n} < \lambda \}}{\lambda} = \frac{C}{4\pi},
\end{equation}
where 
\begin{equation}\label{see}
C= \pi^2 \sum_i \frac{1}{\sqrt{\mu_i\xi_i}} \leq \Area(\scrD).
\end{equation} 
This may either be verified directly, or concluded from the observation (cf. Sections \ref{secProof} and \ref{secProof2}) that $\{\mu_{i,m,n}\}$ can be identified with the spectrum of the Dirichlet Laplacian on an infinite union of rectangles $\scrD_i$ with sides $\ell_i=\pi\xi_i^{-1/2}$, $\delta_i=\pi\mu_i^{-1/2}$, and thus total area $C=\sum_i\Area(\scrD_i)$. In this interpretation, \eqref{weyl2} represents Weyl's law \eqref{weyl}.
\end{remark}

\begin{remark}
If assumption \eqref{assu} holds e.g. for the quasimodes $\psi_{i,1,1}$, eqs. \eqref{tru} and \eqref{three} imply there is an infinite sequence of eigenfunctions $\varphi_{j_i}$, such that for any compact $\scrK\subset\scrD$
\begin{equation}
\liminf_{i\to\infty} \|\varphi_{j_i}\|_{\scrD-\scrK} \geq \frac{1-b^{-1}}{\sqrt{k}} .
\end{equation}
That is, the eigenstates $\varphi_{j_i}$ loose a positive proportion of mass. It should be stressed that we have not ruled out the probably very remote possibility that assumption \eqref{assu} with $\epsilon_i=O(m\xi_i)$ can never be satisfied for the domains $\scrD$ considered in the theorem (an explicit construction of $\scrD$ is given in Section \ref{secProof}). It would be interesting to see whether \eqref{assu} can be established at least for generic choices of such $\scrD$, i.e., generic choices of $\delta_i$. 
In Section \ref{secThm2} we will prove an upper bound for the error term in Weyl's law, which in turn yields a rough estimate on possible level clustering.
\end{remark}

\begin{remark}
For $m,n$ bounded as $i\to\infty$ the theorem establishes quasimodes with very small discrepancy,
\begin{equation} \label{two5}
\| (\Delta+\mu_{i,m,n}) \psi_{i,m,n} \| = O\bigg(\frac{h(\mu_{i,m,n})^2}{\mu_{i,m,n}\, \tau(\mu_{i,m,n})^2}\bigg)  \| \psi_{i,m,n} \| .
\end{equation}
Since $h$ and $\tau$ can be arbitrarily slowly increasing/decreasing functions (respectively), this yields quasimodes of order arbitrarily close to 2; cf. example \ref{exalg} below. The number of such quasimodes with $\mu_{i,m,n}<\lambda$,
\begin{equation}
\begin{split}
N_{\text{bb}}(\lambda) & = \#\{ (i,m,n) :\, m,n=O(1),\; \mu_{i,m,n}<\lambda \}\\
& \asymp \#\{ i:\; \mu_i<\lambda \},
\end{split}
\end{equation}
is determined by the restriction that
\begin{equation}
\int \tau(\lambda) dN_{\text{bb}}(\lambda) <\infty.
\end{equation}
Hence the higher the desired accuracy of quasimodes (achieved by choosing a sufficiently slowly decreasing $\tau$), the thinner the corresponding sequence of quasimodes becomes.
\end{remark}

\begin{remark}
The theorem also implies that there can be sequences of quasimodes of order zero that have almost full density. `Order zero' means that
\begin{equation} \label{two6}
\| (\Delta+\mu_{i,m,n}) \psi_{i,m,n} \| = O(1) 
\| \psi_{i,m,n} \|,
\end{equation}
i.e., $m\xi_i \leq C_1$ for some constant $C_1>0$. Since in view of \eqref{eps} there is a constant $C_2>0$ such that $\xi_i\mu_i \geq C_2$, we have
\begin{equation}\label{low}
\begin{split}
N_{\text{BB}}(\lambda) & = \#\{ (i,m,n) :\, \mu_{i,m,n}= n^2 \mu_i + m^2 \xi_i <\lambda, \; m\xi_i \leq C_1 \} \\
& \geq \#\left\{ (i,m,n) :\, n^2<\frac{\lambda}{\mu_i} - \frac{C_1^2}{C_2} , \; m \leq \frac{C_1}{\xi_i} \right\} \\
& \asymp \sqrt\lambda \sum_{\mu_i<\lambda}  \frac{\sqrt{\mu_i}\,\tau(\mu_i)^2}{h(\mu_i)^2}.
\end{split}
\end{equation}
For suitable choices of $h$ and $\tau$ this quantity can be arbitrarily close to a function $\asymp\lambda$, cf. \eqref{low2}. On the other hand, it is bounded from below by $\gg\sqrt\lambda$. This bound is attained in the case when
\begin{equation}
\sum_{i=1}^\infty  \frac{\sqrt{\mu_i}\,\tau(\mu_i)^2}{h(\mu_i)^2} < \infty,
\end{equation}
and coincides with the bound for compact domains, cf. \cite{Donnelly03}. Note that the heuristic approaches in \cite{Backer97,Tanner97} predict a greater number of bouncing ball modes. 
\end{remark}

\begin{ex}\label{exalg}
Take $h(x)=x^{\beta}$ with $0\leq\beta<1/2$.
For any given infinite sequence of real numbers $\mu_i$ with
\begin{equation}\label{one2}
\#\{ j : \mu_j \leq \lambda \} \asymp \lambda^{\alpha} ,
\end{equation}
there is a domain $\scrD$ with $\int f(x)^{1-2\beta} dx < \infty$, so that the corresponding quasimodes $\psi_j$ have order $2-2\sigma$, for any fixed $\sigma>2(\alpha+\beta)$. That is,
\begin{equation} \label{two1}
\| (\Delta+\mu_{i,m,n}) \psi_{i,m,n} \| =O(m \mu_{i,m,n}^{-1+\sigma}) 
\| \psi_{i,m,n} \|,
\end{equation}
The fact that \eqref{one2} implies \eqref{one} with $\tau(x)=x^{-\alpha'}$ ($\alpha'>\alpha$) is seen by summation by parts. In view of Weyl's law \eqref{weyl} and the small discrepancy $O(\mu_{i,m,n}^{-1+\sigma})$ for bounded $m$, a failure of assumption \eqref{assu} would imply an extreme clustering of eigenvalues. As we shall see in Section \ref{secThm2}, the bounds on the error term in Weyl'a law worsen as $\sigma\to 0$, and hence clustering cannot be ruled out.

An evaluation of the lower bound for the number of order-zero quasimodes in \eqref{low} yields
\begin{equation}\label{low2}
N_{\text{BB}}(\lambda) \gg \lambda^\theta,
\end{equation}
with $\theta=\max\{1+\alpha-2\alpha'-2\beta,1/2\}$.
Note that $\theta$ can be arbitrarily close to 1 for suitable parameter choices.
\end{ex}

\begin{ex}
A second interesting choice that yields a domain $\scrD$ with exponentially narrow cusps, is $h(x)=\sqrt{x}/\log^\gamma(1+x)$ with $\gamma>0$.
For any given infinite sequence of real numbers $\mu_i$ with
\begin{equation}\label{one3}
\#\{ j : \mu_j \leq \lambda \} \asymp \log^\alpha\lambda ,
\end{equation}
there is a domain $\scrD$ with $\int |\log f(x)|^{-\gamma}dx < \infty$, so that
\begin{equation} \label{two2}
\| (\Delta+\mu_{i,m,n}) \psi_{i,m,n} \| =O(m\log^{-\sigma}\mu_{i,m,n}) 
\| \psi_i \|,
\end{equation}
for any fixed $\sigma<2(\gamma-\alpha)$. Choose here $\tau(x)=\log^{-\alpha'}x$ with $\alpha'>\alpha$, and \eqref{one} can again be checked using summation by parts.

In this case the number of order-zero quasimodes is bounded from below by
\begin{equation}\label{low3}
N_{\text{BB}}(\lambda) \gg \sqrt\lambda.
\end{equation}
\end{ex}

\section{Proof of Theorem \ref{thm1}\label{secProof}}

We begin by constructing accurate quasimodes on the rectangle
$[a,a+\ell] \times [0,\delta]$ with Dirichlet boundary conditions at $y=0,\delta$.
Let $\chi\in C_0^\infty(\RR)$ be a mollified characteristic function of the interval $[0,1]$. That is, $0\leq\chi(x)\leq 1$, $\chi(x)=0$ for $x\notin[0,1]$ and $\chi(x)=1$ for $x\in[\epsilon,1-\epsilon]$ for some fixed, small $\epsilon>0$. We assume also that $\chi'(x)=O(\epsilon^{-1})$ (such a choice is always possible).
For $m,n\in\NN$, $a\in\RR$ and $\ell,\delta>0$ put
\begin{equation}\label{qm}
\psi_{m,n}(x,y)= \chi\bigg(\frac{x-a}{\ell}\bigg) \sin\bigg(\frac{\pi m (x-a)}{\ell}\bigg) \sin\bigg(\frac{\pi n y}{\delta}\bigg)
\end{equation}
and
\begin{equation}\label{qmu}
\mu_{m,n} = \pi^2 \bigg[ \bigg(\frac{m}{\ell}\bigg)^2 + \bigg(\frac{n}{\delta}\bigg)^2 \bigg].
\end{equation}
Straightforward differentiation yields
\begin{multline}
(\Delta+\mu_{m,n}) \psi_{m,n}(x,y)
= \frac{1}{\ell^2} \bigg[ 2\pi m \chi'\bigg(\frac{x-a}{\ell}\bigg) \cos\bigg(\frac{\pi m (x-a)}{\ell}\bigg) \\ + \chi''\bigg(\frac{x-a}{\ell}\bigg) \sin\bigg(\frac{\pi m (x-a)}{\ell}\bigg) \bigg] \sin\bigg(\frac{\pi n y}{\delta}\bigg),
\end{multline}
and hence
\begin{equation}
\| (\Delta+\mu_{m,n}) \psi_{m,n} \|^2 = O_\chi\bigg(\frac{m^2\delta}{\ell^3}\bigg).
\end{equation}
where the implied constant only depends on the choice of $\chi$. 
Because of this and
\begin{equation}
\| \psi_{m,n} \|^2 = \frac{\ell\delta}{4} (1+O(\epsilon)),
\end{equation}
we obtain
\begin{equation}\label{err}
\| (\Delta+\mu_{m,n}) \psi_{m,n} \| =  O_\chi\bigg(\frac{m}{\ell^2}\bigg) 
\| \psi_{m,n} \| .
\end{equation}

Furthermore, for $n\neq n'$ we have $\langle \psi_{m,n},\psi_{m',n'} \rangle = 0$, and for $n=n'$, $m\neq m'$,
\begin{equation}
\begin{split}
\langle \psi_{m,n},\psi_{m',n} \rangle 
& = \frac{\delta}{2} \int_0^{\ell} \chi\bigg(\frac{x}{\ell}\bigg)^2 \sin\bigg(\frac{\pi m x}{\ell}\bigg)
\sin\bigg(\frac{\pi m' x}{\ell}\bigg) dx \\
& = \frac{\delta}{2} \bigg\{ \int_0^{\epsilon\ell} + \int_{(1-\epsilon)\ell}^\ell \bigg\} \bigg[\chi\bigg(\frac{x}{\ell}\bigg)^2-1\bigg] \sin\bigg(\frac{\pi m x}{\ell}\bigg)
\sin\bigg(\frac{\pi m' x}{\ell}\bigg)dx \\
&  = \frac{\ell\delta}{4} \bigg\{ \int_0^{\epsilon} + \int_{1-\epsilon}^1 \bigg\} [\chi(x)^2-1] [\cos(\pi (m-m') x) -\cos(\pi (m+m') x)] dx\\
& = \frac{\ell\delta}{4} O(\epsilon).
\end{split}
\end{equation}
On the other hand, using integration by parts, we have
\begin{multline}
\int_0^{\epsilon} [\chi(x)^2-1] \cos(\pi (m-m') x) dx \\
=
\frac{1}{\pi(m-m')} \bigg\{ \bigg[ [\chi(x)^2-1] \sin(\pi (m-m') x) \bigg]_0^\epsilon
\\ - \int_0^{\epsilon} 2\chi(x)\chi'(x) \sin(\pi (m-m') x) dx \bigg\}. 
\end{multline}
Since $\chi(\epsilon)^2=1$, $\sin(0)=0$ the first term vanishes, and since $\chi'(x)=O(\epsilon^{-1})$ the integral is of $O(1)$. The analogous argument works for the remaining integrals. Hence
\begin{equation}\label{2b}
|\langle \psi_{m,n},\psi_{m',n} \rangle| \ll \min\bigg\{\epsilon,\frac{1}{|m-m'|}\bigg\}
\|\psi_{m,n}\|\,\|\psi_{m',n}\|.
\end{equation}

We will now give an explicit construction of $\scrD$. The function $f$ is chosen constant on the intervals $[a_i,a_{i+1})$, $i=1,2,3,\ldots$; set $\delta_i=f(a_i)$ and $\ell_i=a_{i+1}-a_i$. As quasimodes we take 
\begin{equation}
\psi_{i,m,n}(x,y)= \chi\bigg(\frac{x-a_i}{\ell_i}\bigg) \sin\bigg(\frac{\pi m(x-a_i)}{\ell_i}\bigg) \sin\bigg(\frac{\pi n y}{\delta_i}\bigg) ,
\end{equation}
with quasi-eigenvalues
\begin{equation}\label{qmu2}
\mu_{i,m,n} = \pi^2 \bigg[ \bigg(\frac{m}{\ell_i}\bigg)^2 + \bigg(\frac{n}{\delta_i}\bigg)^2 \bigg].
\end{equation}
By construction, these are completely localized in the rectangle $[a_i,a_{i+1}]\times[0,\delta_i]$ and hence satisfy requirement (iv) of the theorem. Setting $\mu_i = \pi^2 \delta_i^{-2}$, every given sequence of $\mu_i$ having property \eqref{one} determines a sequence of $\delta_i$. Because of \eqref{err}, 
\begin{equation}
\frac{\| (\Delta+\mu_{i,m,n}) \psi_{i,m,n} \|}{\| \psi_{i,m,n} \|} =
O_\chi(m \ell_i^{-2})=O_\chi(m \delta_i^2 A_i^{-2})=O_\chi(m \mu_i^{-1} A_i^{-2}).
\end{equation}
To minimize the discrepancy, we would like to choose $A_i$ as large as possible. The choice $A_i=\tau(\mu_i) h(\mu_i)^{-1}$ yields condition (i) and determines $f$. Since
\begin{equation}\label{fina1}
\begin{split}
\int_{a_1}^\infty  f(x) h(\pi^2 f(x)^{-2}) dx & = \sum_i \ell_i  f(a_i) h(\pi^2 f(a_i)^{-2})\\ 
& = \sum_i A_i h(\pi^2 \delta_i^{-2}) \\
& = \sum_i \tau(\mu_i)  < \infty  ,
\end{split}
\end{equation}
the function $f$ is in the required class satisfying \eqref{fina}. 

Condition (ii) is evident from \eqref{qm}, and (iii) from \eqref{2b}. 

\section{Asymptotic distribution of eigenvalues\label{secThm2}}

In view of condition \eqref{assu} we would like to control the number of eigenvalues in small intervals. The following theorem illustrates that extreme level clustering cannot a priori be ruled out.

\begin{thm}\label{thm2}
The spectral counting function $N(\lambda)=\#\{j:\lambda_j < \lambda\}$ of the Dirichlet Laplacian for the domain $\scrD$ (as in Section \ref{secProof}) satisfies
\begin{equation}
N(\lambda) = \frac{\Area(\scrD)}{4\pi}\,\lambda - \frac{L(\lambda)}{4\pi} \sqrt\lambda + \frac{1}{2\pi} \sqrt\lambda
\sum_{\substack{i=1\\ \delta_i\sqrt\lambda>\pi}}^\infty \ell_i  \sum_{r=1}^\infty \frac1r J_1\bigg(2 r \delta_i \sqrt{\lambda}\bigg) +O(\sqrt\lambda),
\end{equation}
where 
\begin{equation}
L(\lambda) = 2 \sum_{\substack{i=1\\ \delta_i\sqrt\lambda>\pi}}^\infty \ell_i 
\end{equation}
and $J_1$ is the $J$-Bessel function.
\end{thm}

\begin{remark}
The standard bound 
\begin{equation}\label{Jbound}
	|J_1(x)|\ll x^{-1/2}\quad \text{for $x$ large}
\end{equation}
implies that
\begin{equation}\label{Neee}
N(\lambda) = \frac{\Area(\scrD)}{4\pi}\,\lambda + O(L(\lambda)\sqrt\lambda),
\end{equation}
where
\begin{equation}
L(\lambda) = 2\pi \sum_{\substack{i=1\\ \mu_i<\lambda}}^\infty \frac{1}{\sqrt{\xi_i}} \ll
\sum_{\substack{i=1\\ \mu_i<\lambda}}^\infty \frac{\sqrt{\mu_i}\,\tau(\mu_i)}{h(\mu_i)} ;
\end{equation}
recall that $\mu_i=\pi^2/\delta_i^2$ and $\xi_i=\pi^2/\ell_i^2$.
As the examples following Theorem \ref{thm1} illustrate, a good quasimode discrepancy ($\xi_i$ small) is thus traded with an error bound in \eqref{Neee} approaching $o(\lambda)$. But as we shall see in the following section, cf. eq. \eqref{DiDi}, the number of eigenvalues in the interval $[\lambda,\lambda+\sigma]$ with $\sigma<\sqrt\lambda$ is
\begin{equation}
N(\lambda+\sigma)-N(\lambda)= \#\{ (i,m,n)\in\NN^3 : \lambda\leq \mu_{i,m,n} < \lambda+\sigma \} +O(\sqrt\lambda), 
\end{equation}
with quasi-eigenvalues $\mu_{i,m,n}$ as in \eqref{quas}.
That is, all extreme fluctuations beyond $O(\sqrt\lambda)$ are due to the presence of bouncing ball quasimodes.
\end{remark}

\section{Proof of Theorem \ref{thm2}\label{secProof2}}

Consider the domains $\scrD_i=\{ (x,y)\in\RR^2 : a_i<x<a_{i+1},\; 0< y < f(x)\}$ where $i=0,1,2,\ldots$ and $a_0=0$. Let $\NDi^{(i)}(\lambda)$ be the spectral counting function for the Dirichlet Laplacian for $\scrD_i$, and $\NNe^{(i)}(\lambda)$ the counting function with Neumann conditions on the boundary lines $x=a_i$ and $x=a_{i+1}$ and Dirichlet conditions on the remaining boundary. Set 
\begin{equation}
\NDi(\lambda)=\sum_{i=0}^\infty \NDi^{(i)}(\lambda), \qquad
\NNe(\lambda)=\sum_{i=0}^\infty \NNe^{(i)}(\lambda).
\end{equation}
It is well known (`Dirichlet-Neumann bracketing' \cite{Berg92a,Berg01}) that 
\begin{equation}
\NDi(\lambda) \leq N(\lambda) \leq \NNe(\lambda).	
\end{equation}
For $i=0$ the general error estimate in Weyl's law for compact domains yields
\begin{equation}
\NDi^{(0)}(\lambda) = \frac{\Area(\scrD_0)}{4\pi} \lambda + O(\sqrt\lambda),
\qquad
\NNe^{(0)}(\lambda) = \frac{\Area(\scrD_0)}{4\pi} \lambda + O(\sqrt\lambda).
\end{equation}
For the remaining domains we have
\begin{equation}
\NDi^\Box(\lambda):=\sum_{i=1}^\infty \NDi^{(i)}(\lambda) =\#\{ (m,n,i)\in\NN^3 : n^2 \mu_i + m^2 \xi_i <\lambda \} 
\end{equation}
and
\begin{equation}
\NNe^\Box(\lambda):=\sum_{i=1}^\infty \NNe^{(i)}(\lambda) =
\NDi^\Box(\lambda) + \#\{ (n,i)\in\NN^2 : n^2 \mu_i <\lambda \} .
\end{equation}
Note that
\begin{equation}\label{sixsix}
\NNe^\Box(\lambda)-\NDi^\Box(\lambda) \leq 
\sum_{\mu_i<\lambda} \sqrt{\frac{\lambda}{\mu_i}} = O(\sqrt\lambda) 
\end{equation}
since $\sum_i \mu_i^{-1/2} <\infty$, cf. \eqref{see}. Therefore
\begin{equation}\label{DiDi}
N(\lambda) = \frac{\Area(\scrD_0)}{4\pi} \lambda + \NDi^\Box+O(\sqrt{\lambda}) .
\end{equation}
Now
\begin{equation}
\NDi^\Box(\lambda) = \sum_{\substack{i,n=1\\ n^2\mu_i<\lambda}}^\infty \left[\sqrt{\frac{\lambda-n^2\mu_i}{\xi_i}} +O(1) \right] 
= \sum_{\substack{i,n=1\\ n^2\mu_i<\lambda}}^\infty \sqrt{\frac{\lambda-n^2\mu_i}{\xi_i}} +O(\sqrt\lambda),
\end{equation}
recall the argument in \eqref{sixsix}.
The main term is
\begin{equation}\label{maint}
\begin{split}
\sum_{\substack{i,n=1\\ n^2\mu_i<\lambda}}^\infty \sqrt{\frac{\lambda-n^2\mu_i}{\xi_i}}
& =\sqrt\lambda \sum_{\substack{i=1\\ \mu_i<\lambda}}^\infty \sum_{n=1}^\infty\frac{1}{\sqrt{\xi_i}}   F\bigg(n\sqrt{\frac{\mu_i}{\lambda}}\bigg) \\
& =\frac12 \sqrt\lambda \sum_{\substack{i=1\\ \mu_i<\lambda}}^\infty \sum_{n=-\infty}^\infty\frac{1}{\sqrt{\xi_i}}   F\bigg(n\sqrt{\frac{\mu_i}{\lambda}}\bigg) -\frac12 \sqrt{\lambda} \sum_{\mu_i<\lambda} \frac{1}{\sqrt{\xi_i}}  
\end{split}
\end{equation}
where $F(x)=\sqrt{\max\{ 1-x^2,0 \}}$. The Poisson summation formula yields for the sum over $n$
\begin{equation}\label{series}
\sum_{n=-\infty}^\infty   F\bigg(n\sqrt{\frac{\mu_i}{\lambda}}\bigg)
=
\sqrt{\frac{\lambda}{\mu_i}} \sum_{r=-\infty}^\infty\widehat F\bigg(r\sqrt{\frac{\lambda}{\mu_i}}\bigg)
\end{equation}
where $\widehat F(0)=\pi/2$ and for $y\neq 0$
\begin{equation}
\begin{split}
\widehat F(y) & = \int_{-1}^1 \sqrt{1-x^2}\, \cos(2\pi xy) dx	\\
& = \frac{1}{2y}\, J_1(2\pi y) .
\end{split}
\end{equation}
So
\begin{equation}\label{series2}
\sum_{n=-\infty}^\infty   F\bigg(n\sqrt{\frac{\mu_i}{\lambda}}\bigg)
=
\frac{\pi}{2} \sqrt{\frac{\lambda}{\mu_i}}+  \sum_{r=1}^\infty \frac1r  J_1\bigg(2\pi r\sqrt{\frac{\lambda}{\mu_i}}\bigg) .
\end{equation}
The bound \eqref{Jbound} proves the convergence of the series on the right hand side of \eqref{series2}. This concludes the proof of Theorem \ref{thm2}.

\section*{Acknowledgments}

I thank M. van den Berg, M. Degli Esposti, J. Keating, M. Lenci, Z. Rudnick and R. Schubert for stimulating discussions.

\end{document}